\documentclass[
    ,final            
  ]
  {aipproc}

\usepackage{epsfig}
\layoutstyle{6x9}

\def\bea{\begin{eqnarray*}}
\def\eea{\end{eqnarray*}}

\def\lQ{\Lambda_{\rm QCD}}

\def\als{\alpha_{\rm s}}

\def\siml{{\ \lower-1.2pt\vbox{\hbox{\rlap{$<$}\lower6pt\vbox{\hbox{$\sim$}}}}\ }}     
\def\simg{{\ \lower-1.2pt\vbox{\hbox{\rlap{$>$}\lower6pt\vbox{\hbox{$\sim$}}}}\ }} 
\newcommand{\MS}{\overline{\rm MS}}

\begin{document}

\title{Open Problems in Heavy Quarkonium Physics}


\author{Antonio Vairo}{
  address={Dipartimento di Fisica dell'Universit\`a di Milano and INFN 
           via Celoria 16, 20133 Milan, Italy}
}

\begin{abstract}
Some recent progress and a personal selection of open
problems in heavy quarkonium physics (spectroscopy, decay and production)
inspired by the activity of the Quarkonium Working Group are reviewed.
\end{abstract}

\maketitle

\section{Introduction}
The wealth and quality of new experimental findings and the theoretical
progress in the construction and use of Effective Field Theories (EFTs) of QCD
are among the reasons of the heavy quarkonium physics {\sl renaissance} witnessed during the last years. 
In order to keep track and make immediately available to a larger community
the progress in the field, experimental and theoretical physicists have
gathered in the last three years to form a Quarkonium Working Group \cite{qwg}. 
The offspring of this collaboration have been three workshops, a school and a newly issued 
CERN Yellow Report \cite{Brambilla:2004wf}. In the following we will review some  
recent results and open problems in heavy quarkonium physics.
We refer the reader to  \cite{Brambilla:2004wf} for exhaustive presentations.

\section{Spectroscopy}
There has been in recent years a renewed interest and a noteworthy progress in
the calculation of the heavy quarkonium spectrum from perturbative QCD. 
The progress comes essentially from three directions:
\begin{itemize}
\item[(1)]{progress in the construction of EFTs of QCD suitable 
to describe non-relativistic bound state systems. These have helped to organize higher-order 
calculations and to factorize high-energy perturbative and low-energy non-perturbative 
contributions. For a recent review we refer to \cite{Brambilla:2004jw}.}
\item[(2)]{fixed order calculations \cite{Titard:1993nn,Pineda:1997hz,Brambilla:1999qa,
Kniehl:1999ud,Brambilla:1999xj,Eiras:2000rh,Hoang:2000fm,Melles:2000dq,Kniehl:2002br,Brambilla:2004wu}.}
\item[(3)]{resummation of large contributions (large logs or large contributions 
associated to renormalon singularities) \cite{Luke:1999kz,Hoang:1998nz,Beneke:1998rk,Pineda:2001ra,
Kniehl:2003ap,Penin:2004xi}.}
\end{itemize}
By definition, a perturbative treatment of the bound state is possible 
if the momentum transfer scale $p$ of the heavy quarks in the bound 
state is much larger than $\lQ$. Two situations may occur under this condition \cite{Brambilla:1999xf}. 
Let us call $E$ the typical kinetic energy of the heavy quark and antiquark
in the centre-of-mass reference frame: in a non-relativistic bound state $p
\sim m v \gg E \sim mv^2$, $v$ being the heavy-quark velocity in the bound state. 
The first situation corresponds to quarkonium states for which $E\simg\lQ$.
Under this circumstance the heavy-quarkonium potential is purely perturbative;
we may call this case Coulombic. In the case $E \gg \lQ$ non-perturbative
contributions are encoded into local condensates, in the case $E \sim \lQ$
into non-local ones (see also \cite{Brambilla:2000am} and references
therein). The second situation corresponds to quarkonium states for which 
$p\gg\lQ\gg E$. Under this circumstance the potential contains a  
perturbative part and short-range non-perturbative contributions. 
We may call quasi Coulombic the case in which non-perturbative contributions 
to the potential turn out to be small and may be treated perturbatively in the
calculation of observables. 

Clearly, perturbative calculations applied to heavy-quarkonium ground states
are on a more solid ground that those to higher resonances, 
and those applied to bottomonium more than those to charmonium,  
since the (quasi-)Coulombic case is more likely to be realized. Perturbative determinations of the 
$\Upsilon(1S)$ and $J/\psi$ masses have been used to extract the $b$ and $c$
masses (see {\sl Masses}). The main uncertainty in these determinations
comes from non-perturbative contributions. Information on their size can come from other ground-state observables, 
like the hyperfine splittings (see {\sl Hyperfine splittings}), and the $B_c$ mass (see {\sl $B_c$ mass}). 
Recent studies of higher quarkonium states in perturbation theory may be found in 
\cite{Titard:1993nn,Titard:1994id,Titard:1994ry,Pineda:1997hz,Brambilla:2001fw,Brambilla:2001qk,Kniehl:2002br}. 
Likely only some of the lowest bottomonium resonances may be treated
consistently as Coulombic or quasi-Coulombic bound states. It is surprising, therefore,  
that some of the gross features of the bottomonium spectrum, like the equal spacing of the radial excitations, 
may be qualitatively reproduced by pure perturbative calculations (see {\sl Higher bottomonium states}).

\begin{table}[th]
\addtolength{\arraycolsep}{0.2cm}
\begin{tabular}{|c|c|c|}
\hline
~~~Ref.~~~  & ~~~~~~order~~~~~~ &  ~~~~~~~~~~~~ $m_b^{\overline{\rm MS}}(m_b^{\overline{\rm MS}})$ (GeV) ~~~~~~~~~~~~ 
\\
\hline
\cite{Beneke:1999fe} & NNLO& $4.24 \pm 0.09$ 
\\ 
\cite{Hoang:1998uv} & NNLO& $4.21 \pm 0.09$
\\
\cite{Pineda:2001zq} & NNLO & $4.210 \pm 0.090 \pm 0.025$
\\
\cite{Brambilla:2001qk}  & NNLO & $4.190 \pm 0.020 \pm 0.025$
\\
\cite{Penin:2002zv} & NNNLO &$4.349 \pm 0.070$
\\
\cite{Lee:2003hh} & NNNLO & $4.20 \pm 0.04$
\\
\cite{Contreras:2003zb} & NNNLO & $4.241 \pm 0.070$ 
\\
\hline
\hline
Ref.  & order &  $m_c^{\overline{\rm MS}}(m_c^{\overline{\rm MS}})$ (GeV)  
\\
\hline
\cite{Brambilla:2001fw}  & NNLO & $1.24 \pm 0.020$ \\
\hline
\end{tabular}
\caption{\label{tableMSmasses} Collection of recently obtained values of 
$m_b^{\overline{\rm MS}}(m_b^{\overline{\rm MS}})$ and  $m_c^{\overline{\rm MS}}(m_c^{\overline{\rm MS}})$
from the $\Upsilon$(1S) and $J/\Psi$ masses.}
\end{table} 

\subsection{Masses}
Table \ref{tableMSmasses} shows some recent determinations of the $b$ and $c$ masses 
from the quarkonium ground-state masses. Finite charm-mass effects have been 
included in the analyses of \cite{Brambilla:2001qk,Lee:2003hh,Contreras:2003zb}. 
With the exception of \cite{Beneke:1999fe} the conversion from the threshold (or pole)
masses to the $\MS$ masses has been performed at three-loop  accuracy.  
All NNNLO analyses are only partially complete, since the full three-loop static
potential has not been calculated yet. Different schemes have been used to implement the leading 
renormalon cancellation. This may explain some of the  differences between 
the results. The most substantial discrepancy, which is between 
the result of \cite{Penin:2002zv} and all the other ones, may be possibly ascribed to 
the use of the on-shell scheme, as well as to the way non-perturbative effects have been implemented.
In \cite{Brambilla:2001fw} the charm $\MS$ mass has been extracted from the $J/\Psi$ mass. 

In the above analyses non-perturbative effects constitute the
major source of uncertainty together with possible effects due to
subleading renormalons.  Indeed, in the situation $E\sim
\lQ$  non-perturbative corrections to the spectrum ($\sim \lQ^3/p^2 \sim
mv^4$) and the subleading renormalon in the mass  ($\sim \lQ^2/m \sim  mv^4$) 
are parametrically of the same order as NNLO corrections in the perturbative expansion. 
As long as non-perturbative effects will not be incorporated in a quantitative manner, 
perturbative calculations beyond NNLO will not improve the determination of the 
heavy-quark masses from the quarkonium system. 

The dominant non-perturbative correction to the mass of a quarkonium state 
$|n\rangle$ of LO binding energy $E_n^{(0)}$ is encoded in the expression 
$$
\delta E_{n}   =  
-i {g^2 \over 9}  \!\! \int_0^\infty \!\!\! dt \,  
 \langle n|  {\bf r} e^{it( E_n^{(0)} - h_o)} {\bf r}  | n \rangle  
\;  \langle {\bf E} (t) \,   {\bf E}  (0)  \rangle(\mu), 
$$
$h_o = p^2/m + \als/(6r)$ being the LO octet Hamiltonian and $\mu$ a 
factorization scale \cite{Brambilla:1999xj,Brambilla:1999xf}. 
A precise determination of this formula would need having an accurate
determination of the chromoelectric correlator. In this respect the available 
data are puzzling. Quenched lattice calculations of gluelump masses 
indicate an inverse correlation length for the correlator $\langle {\bf E} (t) \,   {\bf E}  (0)  \rangle$
of about 1.25 GeV \cite{Foster:1998wu,Bali:2003jq}. This scale is not only
larger than the scale $E$ but also of the same order as the
momentum-transfer scale $p$, if $p$ is identified with $m_b/2 \times 4/3 \times \als(p)$
(e.g. $m_b \approx 4.7$ GeV and $\als(p) \approx 0.4$). This may potentially question the
whole perturbative approach to the ground state of bottomonium.
On the other hand this approach is supported by the fact that it provides a value of the $b$
mass consistent with those obtained from low moments sum rule calculations or
from the $B$ system \cite{El-Khadra:2002wp}. Only a better determination 
of the chromoelectric correlator can solve the conundrum. 
A possible solution could come, for instance, from  
\begin{itemize}
\item{unquenching, if it lowers the value of the inverse correlation length as
  calculations with cooling techniques seem to indicate \cite{D'Elia:1997ne}.}
\item{a particular parametrization of the chromoelectric correlator that 
  sets in the non-perturbative behaviour at a scale $\mu$ 
  lower than the inverse correlation length.}
\end{itemize}

The assumption on the Coulombic behaviour of ground-state quarkonia may be
also tested on other observables, like the $B_c$ mass and the hyperfine splittings.

\subsection{$B_c$ mass}
Table \ref{TabBc} shows some recent determinations of the $B_c$ mass in perturbation theory at NNLO 
accuracy compared with the value of the $B_c$ mass in  \cite{Eidelman:2004wy} and in a very recent 
lattice study \cite{Allison:2004be}. 
The $b$ and $c$ pole masses are expanded in terms of the $\Upsilon(1S)$ and $J/\psi$ masses 
respectively in \cite{Brambilla:2000db}, and in terms of the $\MS$ masses in \cite{Brambilla:2001fw,Brambilla:2001qk}.
In \cite{Brambilla:2001qk} finite charm-mass effects are included. Since the charm quark effectively decouples 
in the scheme of Ref.~\cite{Brambilla:2000db}, we may consider the results of Refs.~\cite{Brambilla:2000db} and 
\cite{Brambilla:2001qk} at the same level of theoretical accuracy and the difference between them as due to higher-order 
corrections beyond NNLO accuracy.

\begin{table}[ht]
\begin{tabular}{|c|c|cccc|}
\hline
\multicolumn{6}{|c|}{$B_c$ mass ~(MeV)}\\
\hline
State &  \cite{Eidelman:2004wy} ~(expt) &~ \cite{Allison:2004be} ~(lattice)~ & ~ \cite{Brambilla:2000db} ~(NNLO)
~& ~\cite{Brambilla:2001fw} ~(NNLO)~& ~\cite{Brambilla:2001qk} ~(NNLO)~\\
\hline
$1^1S_0$ & $6400(400)$ &$6304\pm12^{+12}_{-0}$ & 6326(29) & 6324(22) & 6307(17) 
\\
\hline
\end{tabular}
\caption{Different perturbative determinations of the $B_c$ mass compared with the experimental 
value and a recent lattice determination.}
\label{TabBc}
\end{table}

In {\it Fermilab Today}, December 3, 2004, the CDF collaboration has announced the preliminary results 
of a search for the $B_c$, using decays into a $J/\psi$ and a charged pion. 
Their preliminary result is $M_{B_c} = 6287\pm5$ MeV. This value, if confirmed, would support 
the assumption that non-perturbative contributions to the quarkonium ground state are of the 
same magnitude as NNLO or even NNNLO corrections, which would be consistent with a 
$E\simg\lQ$ power counting.

\subsection{Hyperfine splittings}
Charmonium and bottomonium ground state hyperfine splittings have been recently calculated 
at NLL in \cite{Kniehl:2003ap}. The result in the charmonium case is consistent with 
the experimental value. This, again, supports the assumption that non-perturbative contributions 
for the quarkonium ground state are consistent with a $E\simg\lQ$ power counting. 
In the bottomonium case there are not yet experimental data.

\begin{figure}[thb]
\makebox[-8cm]{\phantom b}
\put(10,5){\epsfxsize=7truecm \epsffile{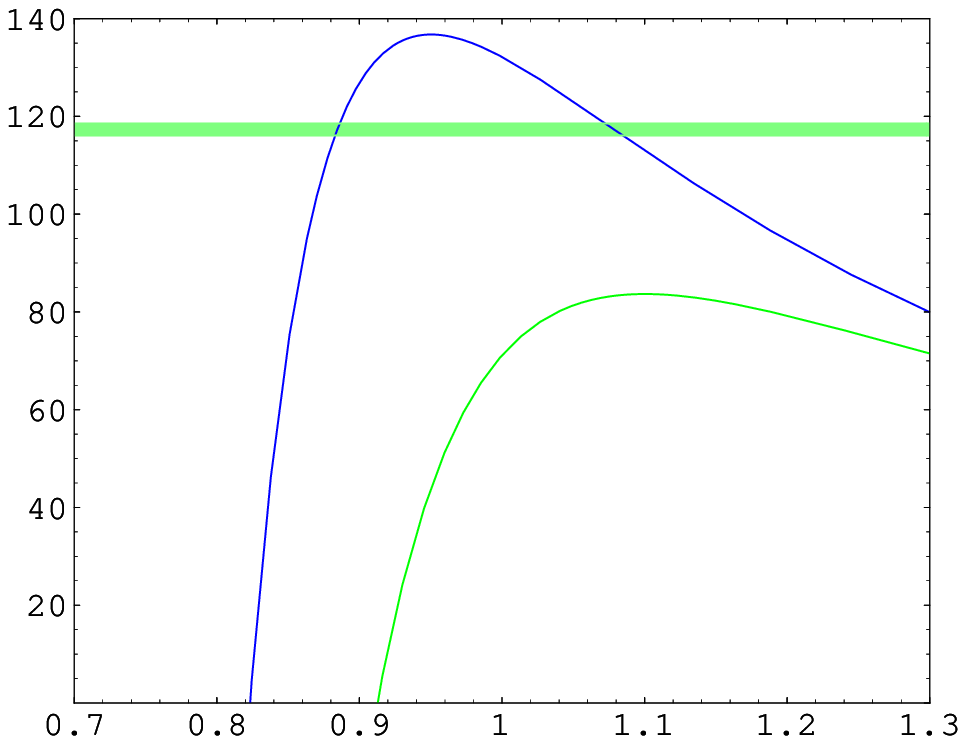}}
\put(250,5){\epsfxsize=7truecm \epsffile{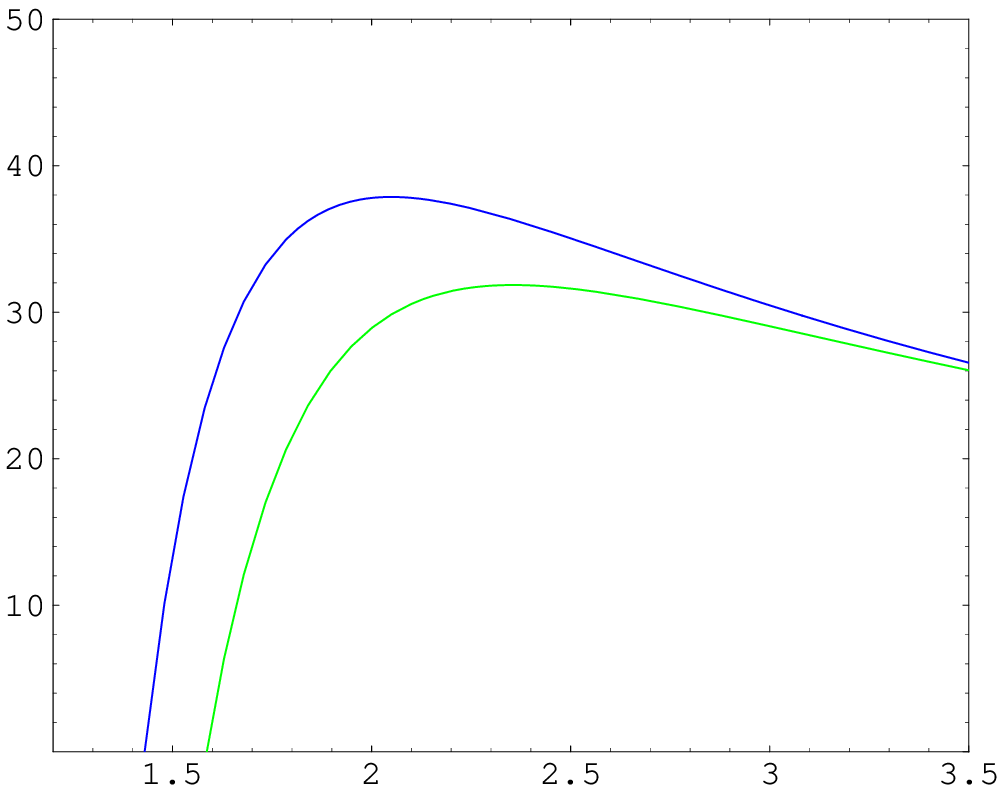}}
\put(-5,145){\small $E_{\rm hfs}$}
\put(-10,130){\small (MeV)}
\put(225,145){\small $E_{\rm hfs}$}
\put(220,130){\small (MeV)}
\put(160,145){\small $(c\bar{c})(1S)$}
\put(400,145){\small $(b\bar{b})(1S)$}
\put(170,-5){\small $\mu$ (GeV)}
\put(410,-5){\small $\mu$ (GeV)}
\caption{Hyperfine splittings of the 1$S$ charmonium (left) and bottomonium (right) 
states at NLO in the 1$S$ mass scheme, $m_c = M_{J/\psi}/2$, $m_b = M_{\Upsilon(1S)}/2$, 
(lower curves) and in the $\MS$ mass scheme, $m_c = m_c^{\overline{\rm
    MS}}(m_c^{\overline{\rm MS}})(1+4 \als/(3\pi))$, 
$m_b = m_b^{\overline{\rm MS}}(m_b^{\overline{\rm MS}})(1+4 \als/(3\pi))$, (upper curves) 
as a function of the normalization scale $\mu$. 
$\als$ runs at 4-loop accuracy with 3 massless flavours. 
Left plot: $m_c^{\overline{\rm MS}}(m_c^{\overline{\rm MS}})$ $=$ $1.28$ GeV, 
$\als(M_{J/\psi}/2)=0.352$,  $\als(m_c^{\overline{\rm MS}}(m_c^{\overline{\rm MS}}))=0.405$. 
Right plot:  $m_b^{\overline{\rm MS}}(m_b^{\overline{\rm MS}}) = 4.22$ GeV, 
$\als(M_{\Upsilon(1S)}/2)=0.209$,  $\als(m_b^{\overline{\rm MS}}(m_b^{\overline{\rm MS}}))=0.218$.}
\label{Figmassdep}
\end{figure}

We note, however, that NNLO corrections may be potentially important for this
observable, also considering that at this order
new type of corrections will appear for the first time:
\begin{itemize}
\item{non-perturbative corrections, which are parametrically of NNLO in the situation $\lQ \sim E$.}
\item{corrections to the mass. In order to show the possible impact of this type of 
corrections, in Fig.~\ref{Figmassdep} we display the hyperfine splittings at
NLO accuracy calculated using $1/2$ of the bottomonium and charmonium vector ground state 
masses and the $\MS$ masses. The difference between the two determinations (taken around 
the maximum in $\mu$) is about 70\% in the charmonium
case and about 30\% in the bottomonium case.}
\end{itemize}

\begin{table}
\begin{tabular}{|c|c|ccc|}
\hline
\multicolumn{5}{|c|}{$b\bar{b}$ states}\\
\hline
 ~~~~State  ~~~~& ~~~~ expt ~~~~&~~~~~~~~~~~\cite{Brambilla:2001fw}~~~~~~~~~~~ & ~~~~~~~~~~~ \cite{Brambilla:2001qk} 
~~~~~~~~~~~ &~~~~~~~~~~~ \cite{Recksiegel:2002za}~~~~~~~~~~ \\
\hline
$1^3S_1$ & 9460 & 9460 & 9460 & 9460   \\
\hline
$1^3P_2$ & 9913 & 9916(59) & 10012(89) & 9956  \\
$1^3P_1$ & 9893 & 9904(67) & 10004(86) & 9938  \\
$1^3P_0$ & 9860 & 9905(56) &  9995(83) & 9915  \\
\hline
$2^3S_2$ & 10023 & 9966(68) & 10084(102) & 10032 \\
\hline
$2^3P_2$ & 10269 &  & 10578(258) & 10270 \\
$2^3P_1$ & 10255 &  & 10564(247) & 10260 \\
$2^3P_0$ & 10232 & 10268 & 10548(239) & 10246 \\
\hline
$3^3S_1$ & 10355 & 10327(208) & 10645(298) & 10315 \\
\hline
\end{tabular}
\caption{
Masses of $b\bar{b}$ states. The result of Ref. \cite{Brambilla:2001fw} 
comes from a full perturbative calculation up to
$O(m\,\alpha_{\rm s}^4)$ without finite charm-mass corrections;
the result of Ref. \cite{Brambilla:2001qk} comes from a full perturbative calculation up to
$O(m\,\alpha_{\rm s}^4)$ including finite charm-mass corrections;
the result of \cite{Recksiegel:2002za} incorporates full corrections up to
$O(m\,\alpha_{\rm s}^4)$ in the individual levels and full corrections up to
$O(m\,\alpha_{\rm s}^5)$ in the fine splittings, includes finite charm-mass
corrections and also depends on some assumptions about the long-range non-perturbative
behaviour of the static potential. 
Numbers without errors are those without explicit or reliable error estimates in the corresponding works.}
\label{Tabhbs}
\end{table}

\subsection{Higher bottomonium states}
Higher bottomonium resonances have been investigated in the framework of
perturbative QCD most recently in \cite{Brambilla:2001fw,Brambilla:2001qk,Recksiegel:2002za}.
The calculation of \cite{Brambilla:2001fw} is accurate at NNLO, that one of
\cite{Brambilla:2001qk} includes finite charm-mass effects, while
\cite{Recksiegel:2002za} is a numerical calculation 
of the spectrum that includes also NLO spin-dependent potentials.
This last calculation also relies on some assumptions about the long-range
behaviour of the static potential. The results are summarized in Table \ref{Tabhbs}.

\begin{figure}[th]
\makebox[-8cm]{\phantom b}
\put(5,5){\epsfxsize=7truecm \epsffile{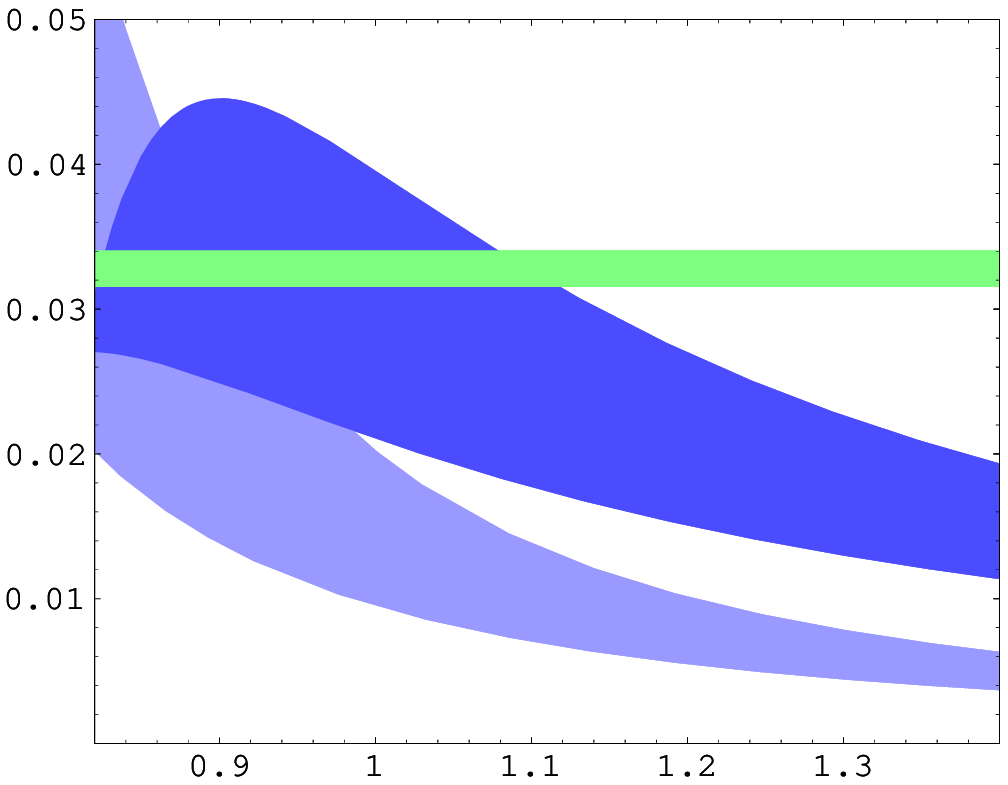}}
\put(240,5){\epsfxsize=7truecm \epsffile{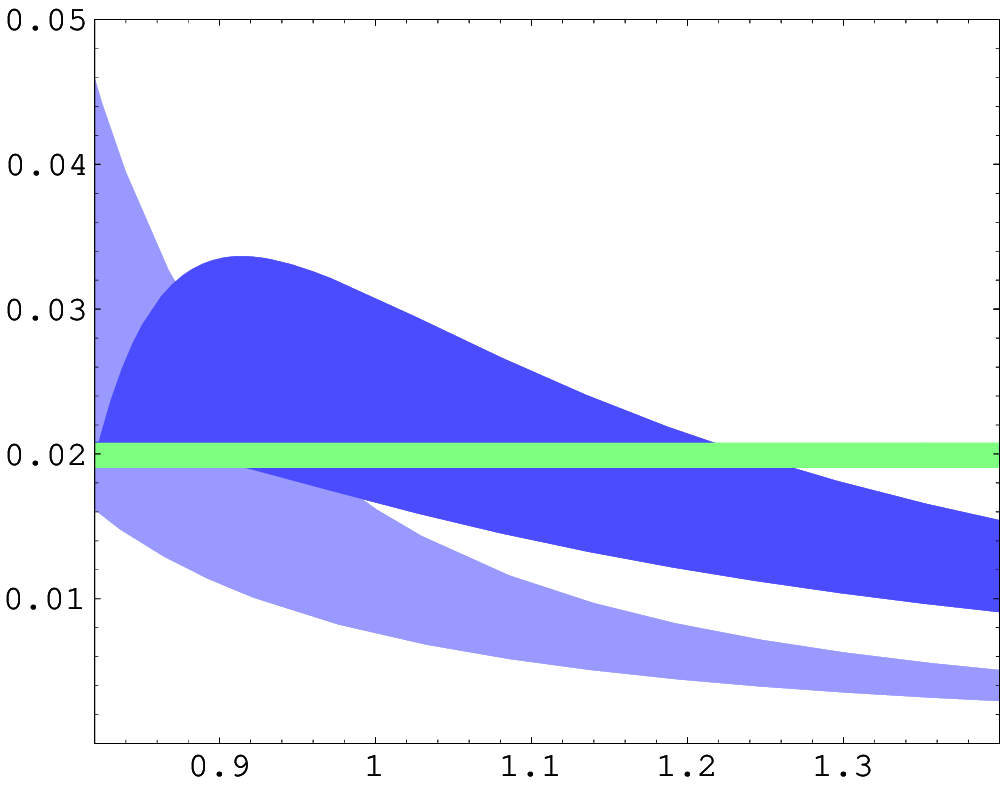}}
\put(115,145){\small $E_{\chi_{b1}(1P)} - E_{\chi_{b0}(1P)}$}
\put(-17,145){\small (GeV)}
\put(350,145){\small $E_{\chi_{b2}(1P)} - E_{\chi_{b1}(1P)}$}
\put(220,145){\small (GeV)}
\put(145,-5){\small $\mu$ (GeV)}
\put(400,-5){\small $\mu$ (GeV)}
\caption{$E_{\chi_{b1}(1P)} - E_{\chi_{b0}(1P)}$ (left) and $E_{\chi_{b2}(1P)}
  - E_{\chi_{b1}(1P)}$ (right) versus the normalization scale $\mu$. 
The light band shows the LO expectation, the dark one the NLO one. 
The widths of the bands account for the uncertainty in $\als(M_Z) = 0.1187 \pm
0.002$ \cite{Eidelman:2004wy}. The horizontal bands show the experimental values \cite{Eidelman:2004wy}. 
From \cite{Brambilla:2004wu}.}
\label{Fig1P}
\end{figure}

\begin{figure}[htb]
\makebox[-4.5truecm]{\phantom b}
\put(20,0){\epsfxsize=7truecm \epsfbox{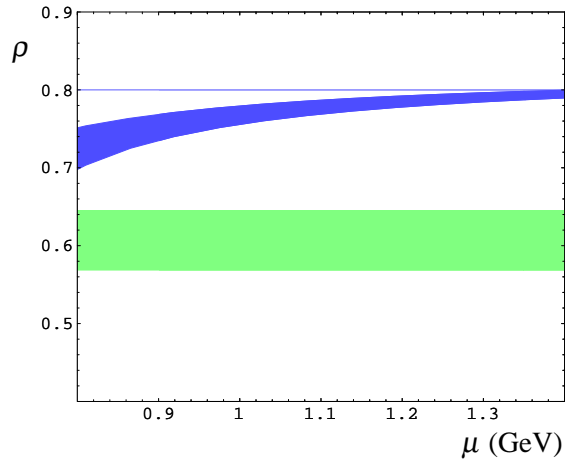}}
\put(180,-10){\small $\mu$ (GeV)}
\put(10,140){\small $\rho$}
\caption{$\rho = (E_{\chi_{b2}} - E_{\chi_{b1}})/(E_{\chi_{b1}} - E_{\chi_{b0}})$ 
versus the normalization scale $\mu$. The horizontal line at 0.8 corresponds 
to the LO expectation, the  curve to the NLO one. The band accounts for the uncertainty in $\als(M_Z) = 
0.1187 \pm 0.002$ \cite{Eidelman:2004wy}. The horizontal band shows the experimental 
value \cite{Eidelman:2004wy}. From \cite{Brambilla:2004wu}.}
\label{figrho}
\end{figure}

The surprising result of these studies is that some gross features of the
lowest part of the bottomonium spectrum, like the approximate equal
spacing of the radial levels, is reproduced by a perturbative calculation
that implements the leading-order renormalon cancellation. If this is coincidental or reflects the
(quasi-)Coulombic nature of the states will be decided by further
studies. A recent NLO calculation of the 1$P$ bottomonium fine splittings 
has been performed in \cite{Brambilla:2004wu}. 
The results are summarized in Fig.~\ref{Fig1P}.
Figure \ref{figrho} plots $\rho$, the ratio of the fine splittings considered
in Fig.~\ref{Fig1P}, as a function of the normalization scale $\mu$
(see \cite{Brambilla:1999ja} and references therein for early studies of this quantity).
It seems to indicate either the existence of large NLL/NNLO
corrections (as it happens in the hyperfine splittings of the 1$S$ levels) 
or sizeable non-perturbative corrections, somehow hidden in the error bands of Fig.~\ref{Fig1P}.

\section{Decay}

\subsection{$\Upsilon$ inclusive decays}
The NRQCD factorization formulas state that inclusive decay widths to light hadrons ($l.h.$)
may be written as sums of products of matrix elements and imaginary
parts of Wilson coefficients of 4-fermion operators \cite{Bodwin:1994jh}. 
In the case of the $\Upsilon$ system they read: 
\bea 
& &\hspace{-5mm} 
 \Gamma( \Upsilon  \to l.h.)
= {2\over m^2}\Bigg(\!\! {\rm Im\,}  f_1(^3 S_1)
 \langle O_1(^3S_1)\rangle_{\Upsilon}
\\
& &\hspace{0mm} 
+ {\rm Im\,} f_{ 8}(^3 S_1) 
 \langle  O_{ 8}(^1S_0) \rangle_{\eta_b}
 + {{\rm Im\,} f_{ 8}(^1 S_0)  \over 3}
 \langle O_{ 8} (^3S_1) \rangle_{\eta_b}
\\
& &\hspace{0mm} 
+ {\rm Im\,}  g_1(^3 S_1)  
{ \langle {\cal P}_1( ^1S_0)\rangle_{\eta_b}  \over m^2} 
 + {\sum_J (2J+1){\rm Im\,}  f_{ 8}(^3 P_J)  \over 9}
{ \langle O_{ 8}(^1P_1)\rangle_{\eta_b}  \over m^2} + \cdots
\Bigg).
\eea

\begin{figure}
\makebox[-8cm]{\phantom b}
\put(5,5){\epsfxsize=7truecm \epsffile{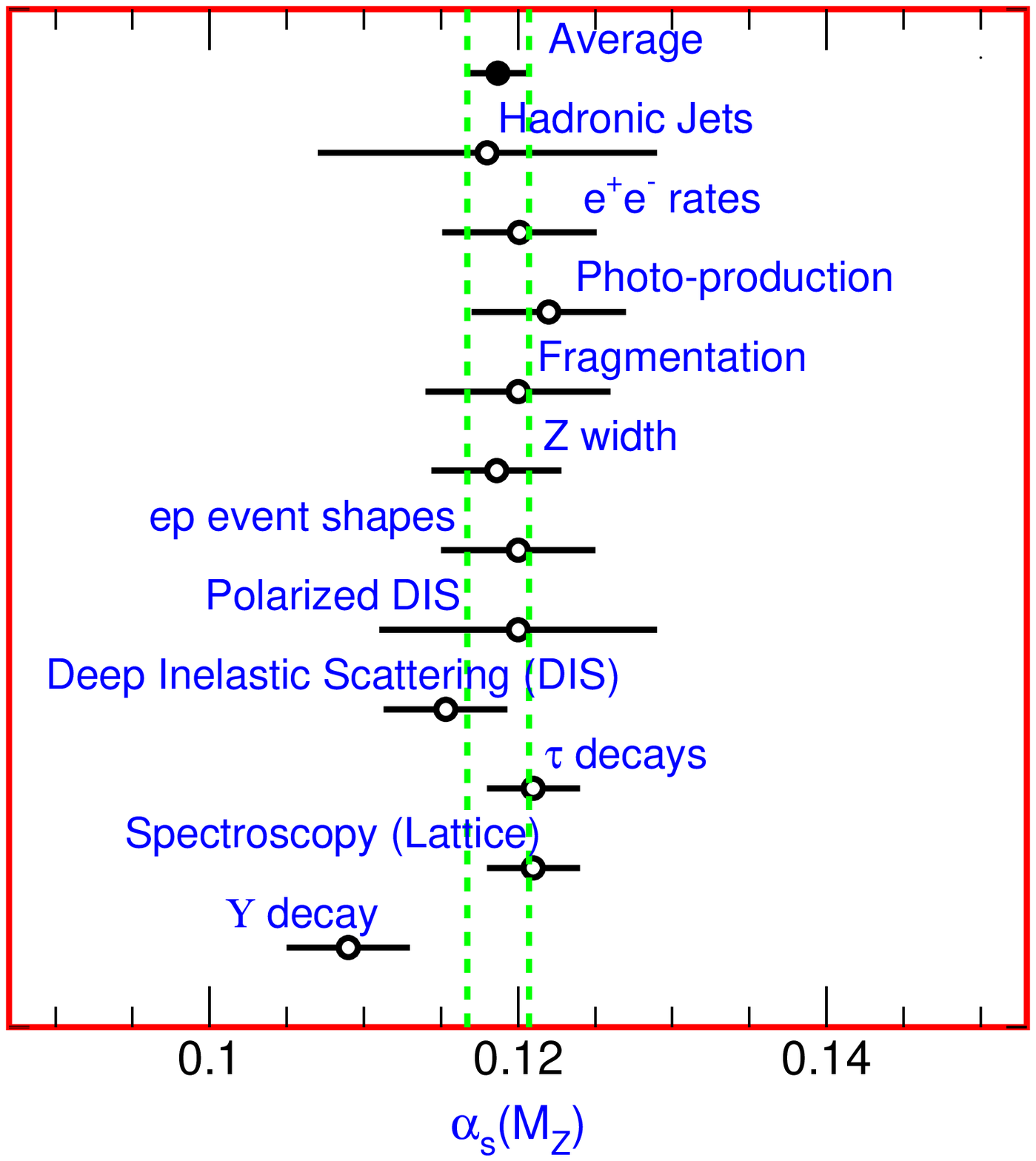}}
\put(240,5){\epsfxsize=7truecm \epsffile{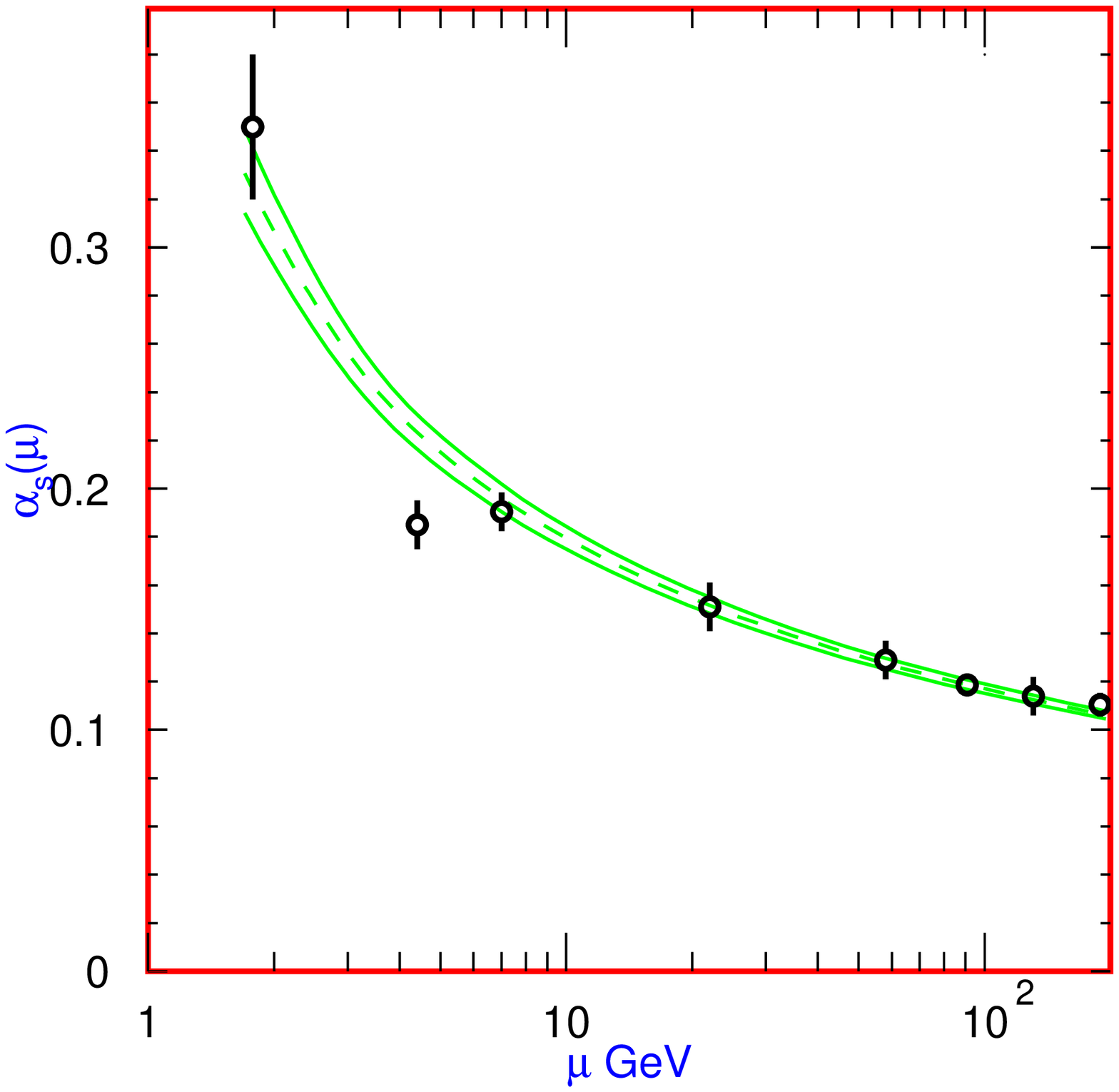}}
\caption{Different determinations of $\als(M_Z)$ (left) and of $\als(\mu)$ at 
different scales $\mu$ (right).  In both figures the determinations outside the band come 
from $\Upsilon$ decay data. From \cite{Eidelman:2004wy}.}
\label{Figalmb}
\end{figure}

The use of these formulas (and similar ones for the electromagnetic decay widths) 
to extract $\als$ at the $b$-mass scale has turned out problematic, as shown by Fig.~\ref{Figalmb}. 
It seems that uncertainties have been underestimated \cite{Brambilla:2004wf}. 
The reasons may be the following:
\begin{itemize}
\item{poor knowledge of the matrix elements. Note that usually octet matrix elements have been 
neglected in the analyses. This may not be justified considering that 
${\rm Im\,}  f_1(^3 S_1)$ and  ${\rm Im\,}  g_1(^3 S_1)$
are suppressed by an extra $\als$ with respect to most of the  ${\rm Im\,}
f_8$ coefficients (see \cite{Vairo:2003gh} and references 
therein) that may compensate, under some circumstances, the suppression in the velocity expansion of the 
octet operator matrix elements. The poor knowledge of the matrix element $\langle  {\cal P}_1  \rangle$ 
has been indicated as one of the major source of uncertainty in \cite{pich02}. 
Studies of decay matrix elements on the lattice can be found in \cite{Bodwin:2004up,Bodwin:2001mk,Bodwin:1996tg}.
Factorization for\-mu\-las for decay matrix elements in pNRQCD have been 
derived in \cite{Brambilla:2001xy,Brambilla:2002nu}.}
\item{large higher-order corrections to the Wilson coefficients \cite{Penin:2004ay}. 
}
\end{itemize}

\begin{table}[th]
\begin{tabular}{|c||c|c||c|c|}
\hline
Ratio &\cite{Eidelman:2004wy} ~(PDG 2004) & \cite{Groom:2000in} ~(PDG 2000) 
&~~ LO~~ &~~ NLO ~~ \\
\hline
 & & & & \\
$\displaystyle {\Gamma(\chi_{c0}\to\gamma\gamma)\over 
\Gamma(\chi_{c2}\to\gamma\gamma)}$ & 5.1$\pm$1.1 &  13$\pm$10  &  $\approx$ 3.75  & $\approx$ 5.43   \\
 & & & & \\
\hline
 & & & & \\
$\displaystyle {\Gamma(\chi_{c2}\to l.h.) - \Gamma(\chi_{c1}\to l.h.) 
\over \Gamma(\chi_{c0}\to\gamma\gamma)}$ 
& 410$\pm$100 &  270$\pm$200  &  $\approx$ 347  & $\approx$ 383   
\\
 & & & & \\
\hline
 & & & & \\
$\displaystyle {\Gamma(\chi_{c0}\to l.h.) - \Gamma(\chi_{c1}\to l.h.) 
\over \Gamma(\chi_{c0}\to\gamma\gamma)}$ 
& 3600$\pm$700 &  3500$\pm$2500  &  $\approx$ 1300  & $\approx$ 
2781   \\
 & & & & \\
\hline
 & & & & \\
$\displaystyle {\Gamma(\chi_{c0}\to l.h.) - \Gamma(\chi_{c2}\to l.h.) 
\over \Gamma(\chi_{c2}\to l.h.) - \Gamma(\chi_{c1}\to l.h.)}$ 
& 7.9$\pm$1.5 &  12.1$\pm$3.2  &  $\approx$ 2.75  & $\approx$ 6.63   \\
 & & & & \\
\hline
 & & & & \\
$\displaystyle {\Gamma(\chi_{c0}\to l.h.) - \Gamma(\chi_{c1}\to l.h.) 
\over \Gamma(\chi_{c2}\to l.h.) - \Gamma(\chi_{c1}\to l.h.)}$ 
& 8.9$\pm$1.1 &  13.1$\pm$3.3  &  $\approx$ 3.75  & $\approx$ 7.63   \\
 & & & & \\
\hline
\end{tabular}
\caption{\label{tab:comp} Comparison of ratios of $\chi_{cJ}$ partial widths.
The experimental values PDG 2004 and PDG 2000 are obtained from the world
averages of \cite{Eidelman:2004wy} and \cite{Groom:2000in} respectively.
The chosen ratios do not depend at leading order in the velocity expansion 
on octet or singlet matrix elements. The LO and NLO columns refer to a leading and  
next-to-leading order calculation done at the renormalization scale $2m_c$ with the following 
choice of parameters: $m_c=1.5$ GeV and $\als(2m_c)=0.245$. From \cite{Brambilla:2004wf}.} 
\end{table}

\subsection{$\chi_c$ inclusive decays}
In the case of $\chi_c$ inclusive decay widths the NRQCD factorization formulas read:
\bea 
&& \hspace{-5mm}  \Gamma(\chi_{cJ}  \to l.h.)
\; = \;
9 \; {\rm Im \,}  f_{ 1}(^3 P_J)  \; 
{ \Big| {{R_{\chi_{cJ}}'}}(0) \Big|^2  \over \pi m^4} 
 \;+\; {2 \; {\rm Im \,}  f_{ 8}(^3 S_1) \over m^2} \; \langle O_{ 8}( {}^1S_0 ) \rangle_{\chi_{cJ}} +\cdots \,,
\\  
\\
&& \hspace{-5mm}  \Gamma ( \chi_{cJ}  \to \gamma \gamma)
\; = \; 
9 \; {\rm Im \,}  f_{\gamma \gamma}(^3 P_J)  \; 
{ \Big| {{R_{\chi_{cJ}}'}}(0) \Big|^2   \over \pi m^4} +\cdots 
\qquad\quad  J=0,2, 
\eea
where $R_{\chi_{cJ}}'(0)$ is the derivative of the radial part of the
$\chi_{cJ}$ wave function at the origin.
The displayed terms are accurate at leading order in the velocity expansion. 
The Wilson coefficients are known at NLO accuracy in $\als$.

The experimental determination of the $\chi_c$ decay widths has dramatically 
improved in the last four years mainly due to the measurements of the E835 
experiment at the Fermilab Antiproton Accumulator. A way to see the impact 
of these measurements is provided by  Tab.~\ref{tab:comp}, where we compare the PDG 2000 \cite{Groom:2000in} with 
the PDG 2004 \cite{Eidelman:2004wy} determinations of different ratios 
of hadronic and electromagnetic widths.  There have been sizable shifts in some central 
values and considerable reductions in the errors. In particular, the error on the ratio 
of the electromagnetic $\chi_{c0}$ and $\chi_{c2}$ widths has been reduced by about a 
factor 10, while in all other ratios the errors have been reduced by a factor 2 or 3.
The considered ratios of hadronic and electromagnetic widths do not depend 
at leading order in the velocity expansion on any non-perturbative parameter. 
Therefore, they can be determined in perturbation theory only.
The last two columns of Tab.~\ref{tab:comp} show the result of a LO and NLO  
calculation respectively. Clearly the data have become sensitive to NLO 
corrections and may be used, in principle, to determine $\als$ at the 
charm-mass scale. Before this, a necessary step is the 
calculation of the decay widths at next-to-leading order in the velocity
expansion, since these contributions are potentially of the same magnitude 
as NLO corrections in the Wilson coefficients  
(see Ref.~\cite{Ma:2002ev} for a calculation in the electromagnetic case).

\section{Production}
In this section we briefly mention two of the main open problems in our understanding of 
charmonium production. We refer to \cite{Brambilla:2004wf} for a proper treatment.

\subsection{Charmonium polarization}
Quarkonium production at large  $p_T$ is dominated by gluon fragmentation.
NRQCD predicts that the dominant gluon-fragmentation process is gluon fragmentation into a 
quark-antiquark pair in a color octet $^3S_1$ state. At large $p_T$ the fragmenting gluon is transversely 
polarized. In the standard NRQCD power counting it is expected that the octet quark-antiquark pair 
keeps the transverse polarization as it evolves into a $S$-wave quarkonium. 
Different power countings \cite{Fleming:2000ib} or higher-order corrections 
may somehow dilute the polarization, which, however, is expected 
to show up in high $p_T$ data. The present Tevatron data
\cite{Affolder:2000nn} do not seem to confirm this expectation.
The uncertainties are, however, too large to make any definite 
claim. The issue here is mainly experimental and more precise Tevatron data are eagerly awaited.

\subsection{Double charmonium production}
The most challenging open problem in charmonium production concerns double charmonium production 
in $e^+e^-$ annihilations. The Belle data of inclusive $J/\psi + c\bar{c}$ production, 
$$
{\sigma(e^+e^-\to J/\psi+ c\bar{c}) \over \sigma(e^+e^-\to J/\psi+ X)} 
= 0.82\pm 0.15\pm 0.14, \quad \hbox{Belle ~\cite{belle-eps2003}},
$$
and of exclusive $J/\psi+\eta_c$ production, 
$$
\sigma(e^+e^-\to J/\psi+\eta_c) \, {\rm Br}(c\bar{c}_{\rm res}\to > \hbox{ 2 ch. track}) 
= 25.6\pm 2.8\pm 3.4~\hbox{fb}, \quad \hbox{Belle ~\cite{Abe:2004ww}}, 
$$
are far by almost an order of magnitude from the available theoretical predictions:
\bea
{\sigma(e^+e^-\to J/\psi+ c\bar{c}) \over \sigma(e^+e^-\to J/\psi+ X)} 
&\approx& 0.1, \quad \hbox{\cite{Cho:1996cg,Baek:1996kq,Yuan:1996ep}},
\\
\sigma(e^+e^-\to J/\psi+\eta_c) 
&=& 2.31\pm 1.09~\hbox{fb}, \quad \hbox{\cite{Braaten:2002fi,Liu:2002wq}}.
\eea
On the other hand, the present upper bound of Belle on $J/\psi+J/\psi$ production is consistent 
with theoretical expectations:
\bea
\sigma(e^+e^-\to J/\psi+J/\psi) \, {\rm Br}(c\bar{c}_{\rm res}\to > \hbox{ 2 ch. track}) 
&<& 9.1~\hbox{fb}, \quad \hbox{Belle ~\cite{Abe:2004ww}}, 
\\
\sigma(e^+e^-\to J/\psi+J/\psi) 
&=& 8.70\pm 2.94~\hbox{fb}, \quad \hbox{\cite{Bodwin:2002fk,Bodwin:2002kk}}.
\eea
Even if independent data by BaBar would be most welcome, the issue here seems 
to be mainly theoretical and the data may be the signal of some new production mechanism.

\end{document}